Figure 1

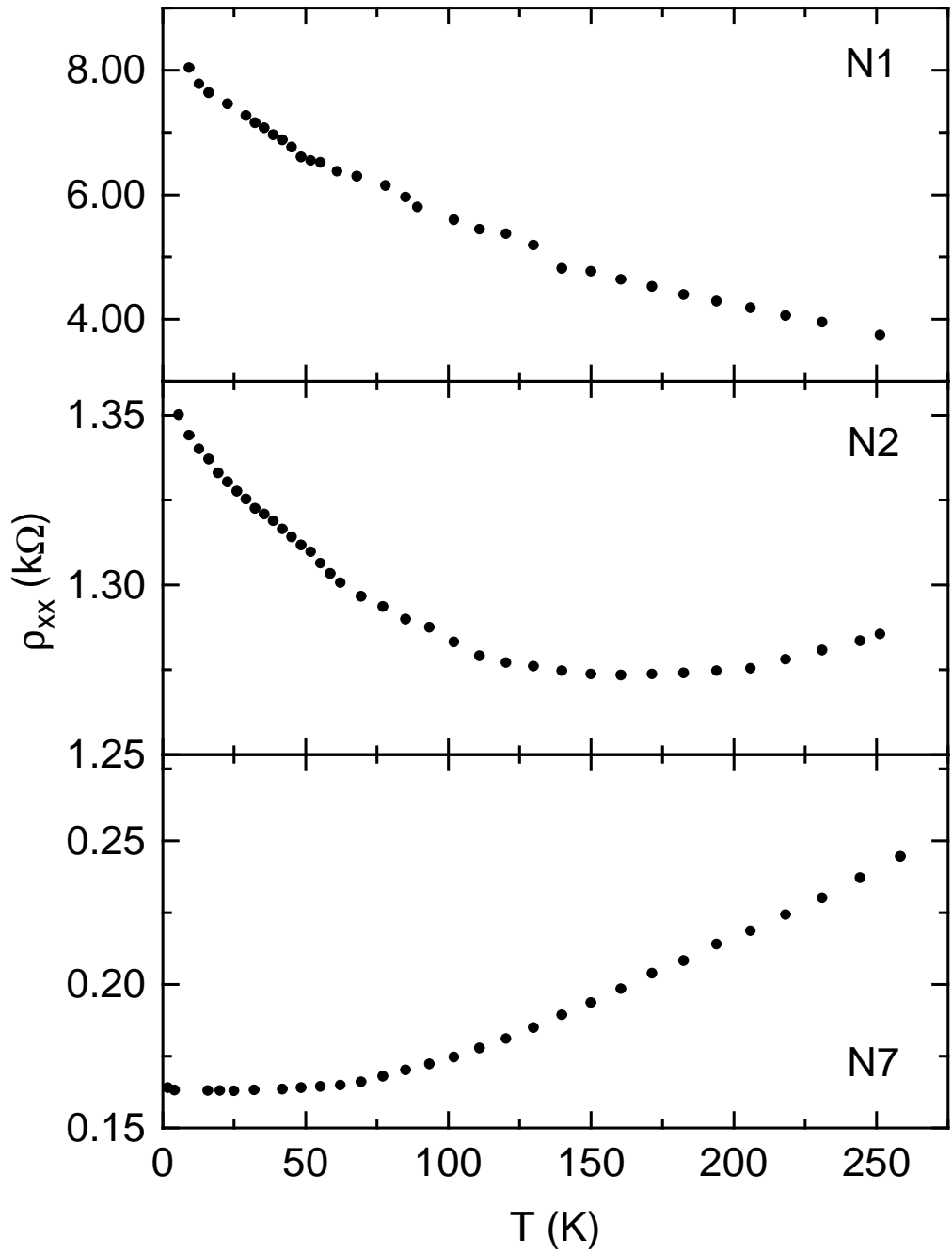

Figure 2.

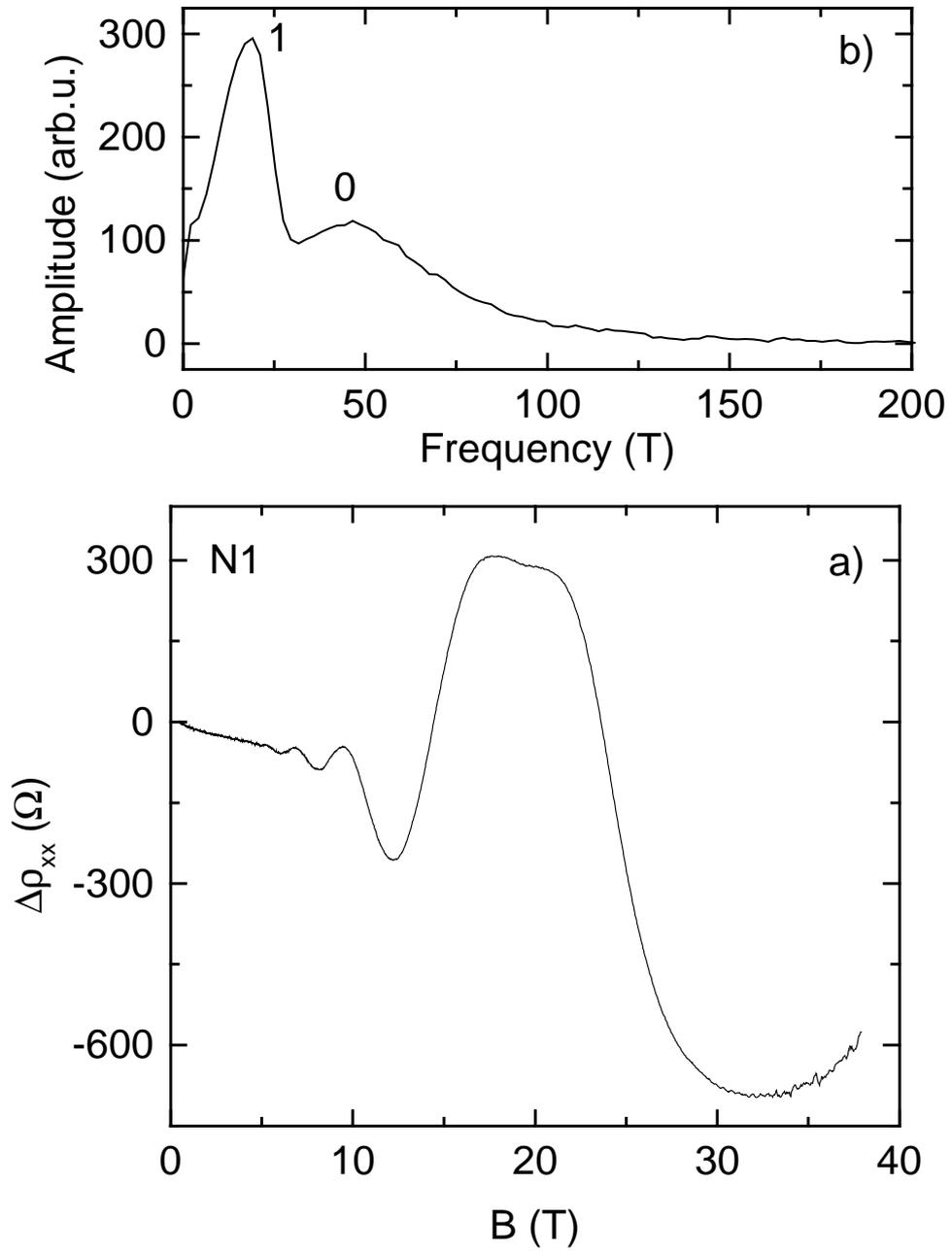

Figure 3.

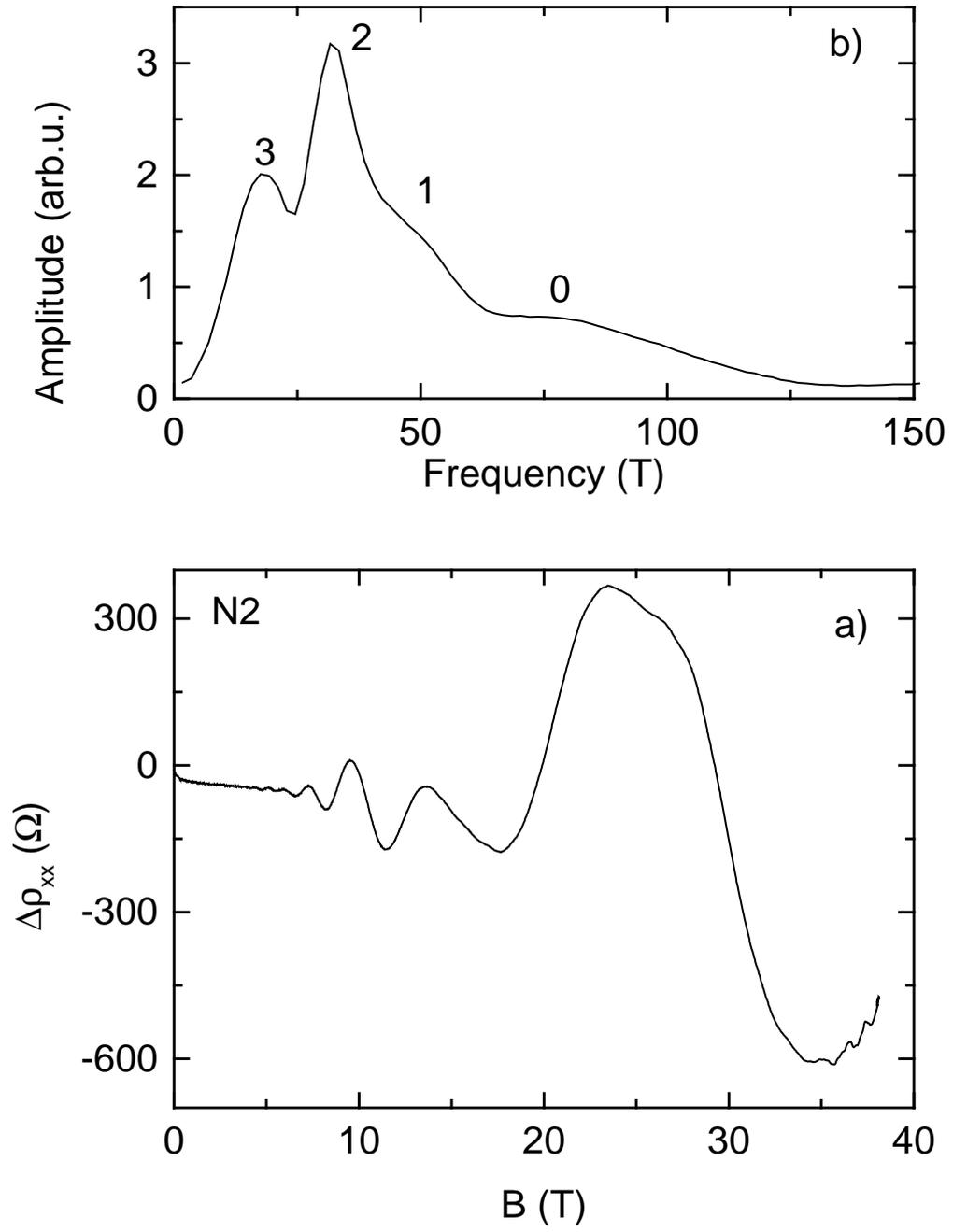

Figure 4.

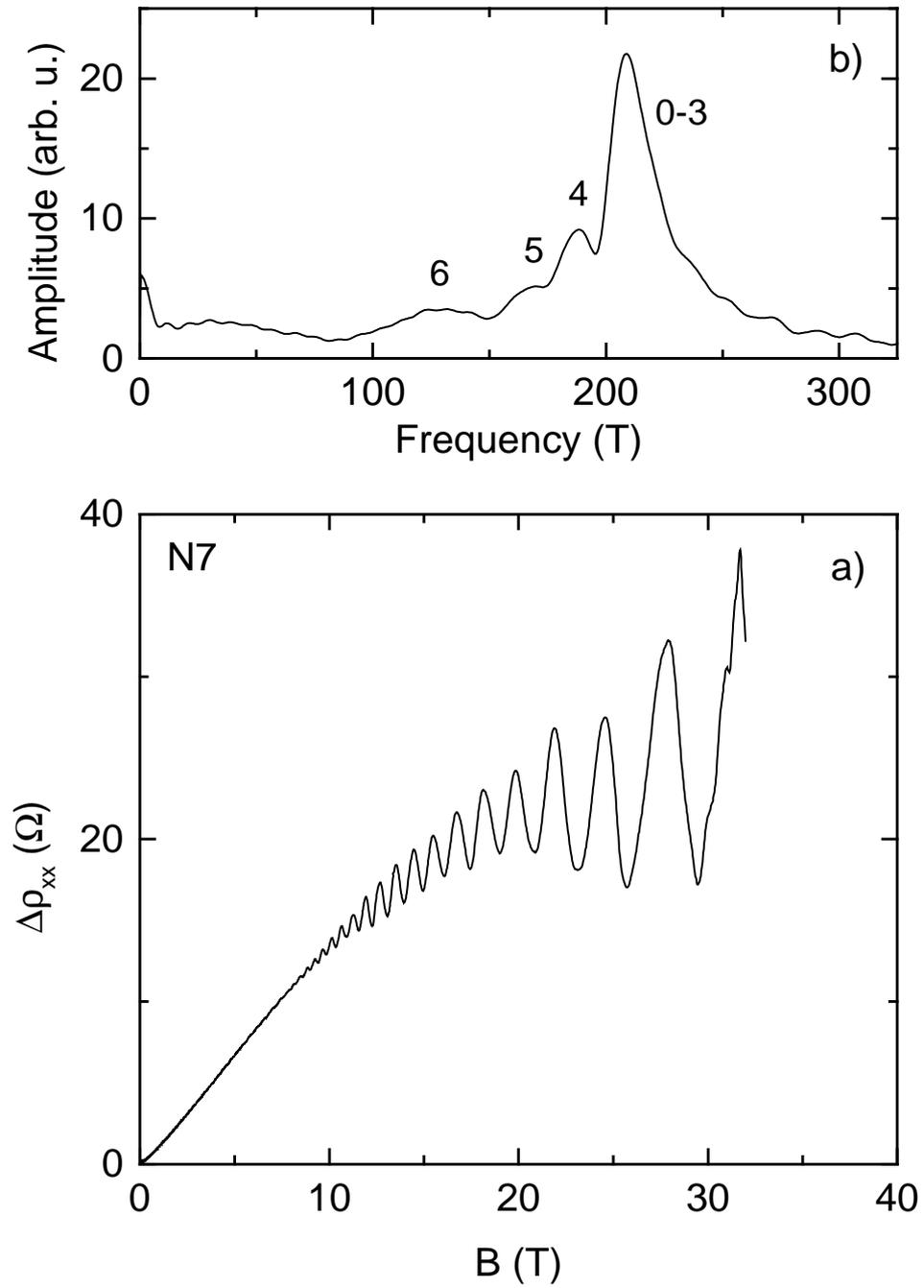

Figure 5.

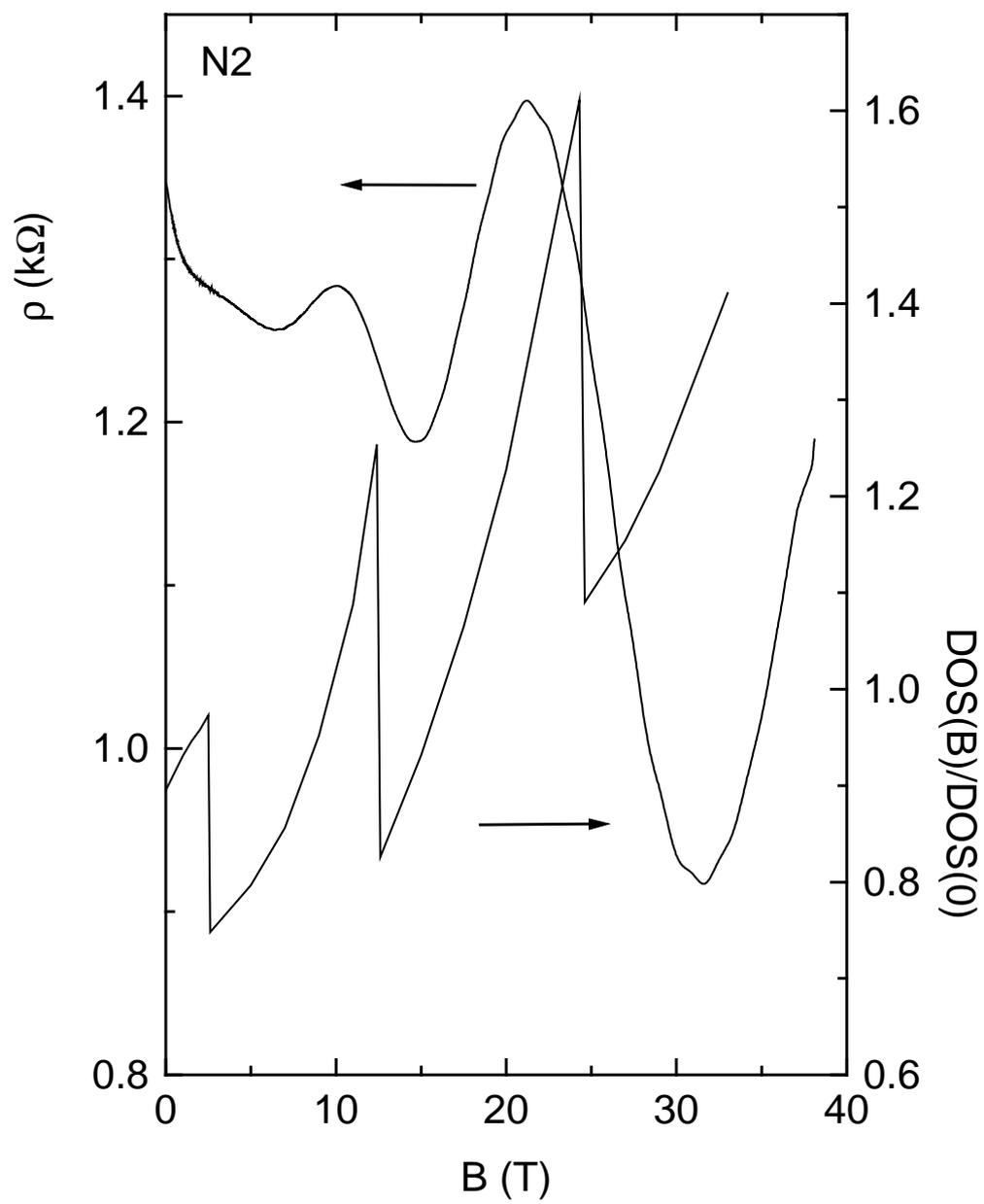

Figure 6.

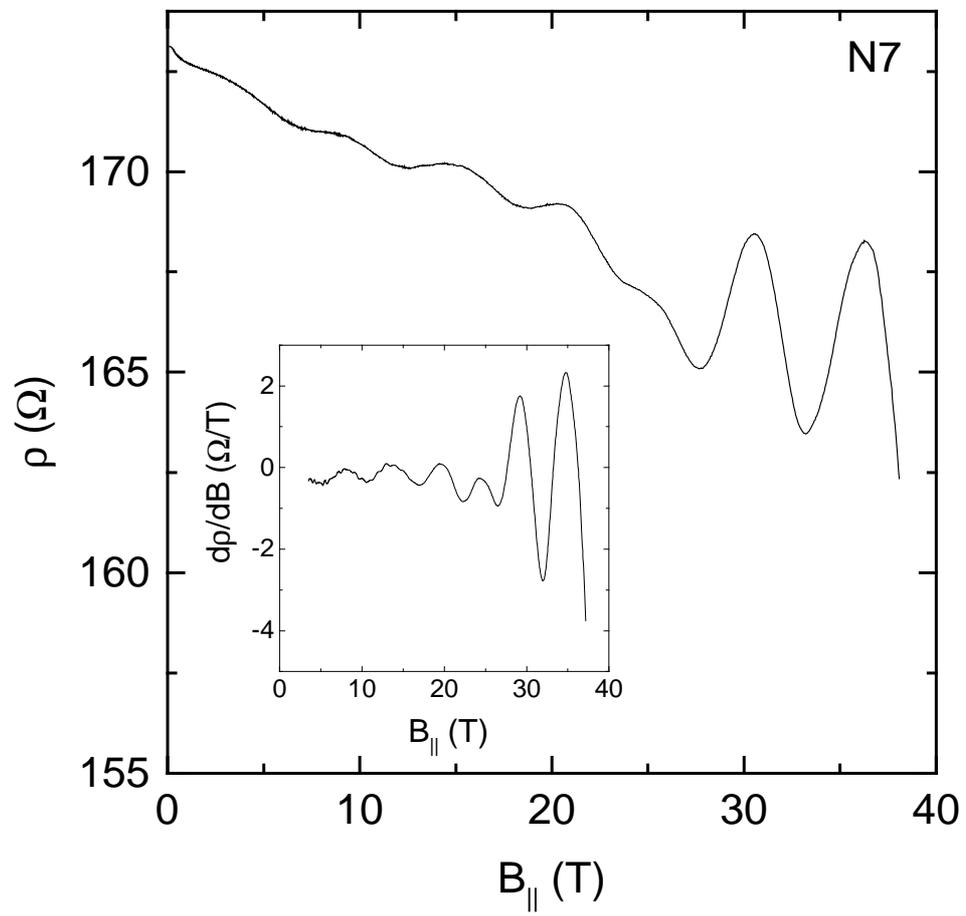

Figure 7.

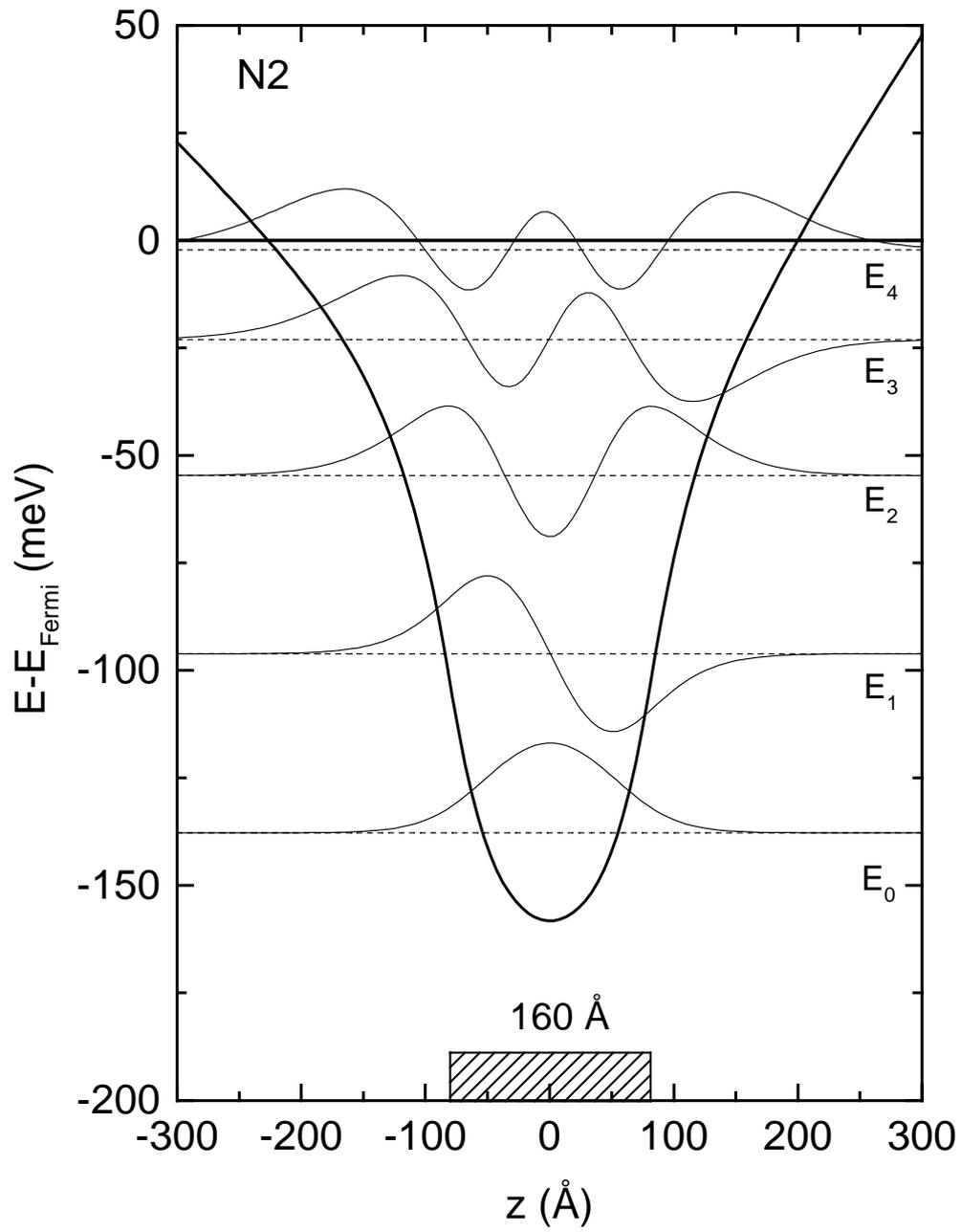

Figure 8

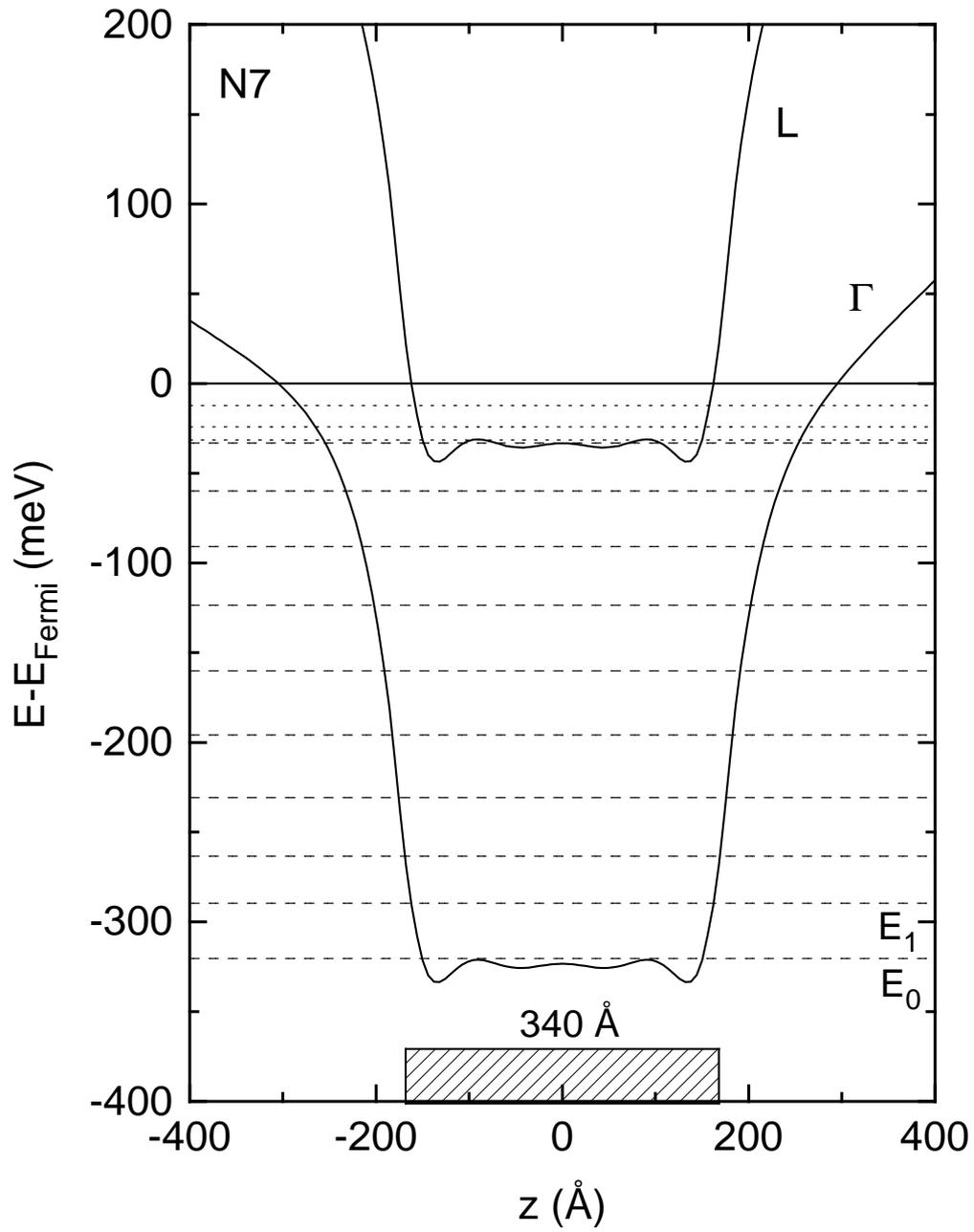

Figure 9.

E−E$_{Fermi}$ (meV)

i=4
i=3
i=2
i=1
i=0

B = 0T
B = 18T

N2

k$_x$ (nm$^{-1}$)

# Sn δ-doping in GaAs

V A Kulbachinskii$, V G Kytin$, R A Lunin$, A V Golikov$, V G Mokerov§, A S Bugaev§, A P Senichkin§, R T F van Schaijk‡, A de Visser‡ and P M Koenraad†

$Low Temperature Physics Department, Moscow State University, 119899, Moscow, Russia
§Institute of Radioengineering and Electronics, Russian Academy of Sciences, 103907 Moscow, Russia
‡Van der Waals-Zeeman Institute, University of Amsterdam, Valckenierstraat 65 1018 XE Amsterdam, The Netherlands
†Physics Department, Eindhoven University of Technology, P.O.Box 513, 5600 MB Eindhoven, The Netherlands

We have prepared a number of GaAs structures δ-doped by Sn using the well-known molecular beam epitaxy growth technique. The samples obtained for a wide range of Sn doping densities were characterised by magnetotransport experiments at low temperatures and in high magnetic fields up to 38 T. Hall-effect and Shubnikov-de Haas measurements show that the electron densities reached are higher than for other δ-dopants, like Si and Be. The maximum carrier density determined by the Hall effect equals $8.4 \times 10^{13}$ cm$^{-2}$. For all samples several Shubnikov-de Haas frequencies were observed, indicating the population of multiple subbands. The depopulation fields of the subbands were determined by measuring the magnetoresistance with the magnetic field in the plane of the δ-layer. The experimental results are in good agreement with selfconsistent bandstructure calculations. These calculation shows that in the sample with the highest electron density also the conduction band at the L point is populated.

Address correspondence to:   R T F van Schaijk
                             Van der Waals-Zeeman Institute
                             Valckenierstraat 65
                             1018 XE Amsterdam
                             the Netherlands
                             Phone: +31 205255795; Fax: +31 205255788
                             E-mail: schaijk@wins.uva.nl

# 1. Introduction

In recent years, interest in the structuring of dopant atoms in semiconductors is growing. Especially, semiconductor structures in which the dopant atoms are confined in a single plane of the host material attract attention, because of promising technological applications. One of the major goals is to engineer structures, which combine high carrier concentrations with high mobilities. Besides, structuring of dopant atoms enables the further miniaturisation of devices, which offers many advantages in applications, notably when high operating speeds are demanded.

A commonly used route to achieve narrow-doping profiles in GaAs is by incorporating mono or δ-doped layers of for instance Si or Be in the structure by molecular beam epitaxy (MBE) growth (see Ref.[1] and references therein). The charge carriers released from the dopant atoms in the δ-layer are confined by the potential well induced by the ionised dopant atoms and form a two-dimensional electron gas (2DEG). The most often used dopant atom in GaAs is silicon, which acts as a n-type dopant. The first observation of a 2DEG in a Si δ-doped layer was reported by Zrenner et al.[2]. Beryllium is most widely used as a p-type dopant.

The structural and electronic properties of Si δ-doped GaAs have been investigated in great detail [1]. Characterisation techniques like SIMS and CV profiling show that the measured thickness of the doping layer is limited by the resolution of the technique. However it can be concluded that dopants can be confined to within ~15 Å, which corresponds to roughly three atomic layers, for samples grown under optimised conditions [3]. Using Si as dopant, fairly large electron densities, $n_e$, of the order of ~$10^{13}$ cm$^{-2}$ can be achieved, while the Hall mobility is typically of the order of $10^3$ cm$^2$/Vs.

In this work, we present the first experimental study of Sn δ-doped GaAs structures. Tin has rarely been used as δ-dopant in GaAs, which is mainly due to its ability to segregate to the surface during the MBE growth process [4]. However, it has been shown that growth at a relatively low substrate temperature (~450 °C) limits the segregation kinetics, which enables the use of Sn for δ-doping. An advantage of the n-type dopant Sn, when compared to Si, is that Sn is less amphoteric, which brings higher carrier concentrations within reach. At the same time, this permits the population of a large number of subbands as usual in Si-δ-doped layers.

The relatively high segregation velocity of Sn may turn into an advantage for growth on misoriented (vicinal) substrates. δ-doping of Sn on vicinal GaAs substrates offers a route to modulate the two dimensional electron gas with a periodic one-dimensional potential. Moreover, by accumulating Sn atoms at the step edges, one could in principle fabricate arrays of quantum wires. Control of the segregation of Sn to the step edges is here the key parameter. Indeed, effects of a reduction of the dimensionality in Sn-doped GaAs structures grown on substrates misoriented at small angles (0.3-3°) from the [100] direction towards the [110] direction (terrace widths of 530-53 Å) have been observed in the magnetotransport properties [5-7].

Our GaAs(δ-Sn) structures have been grown by MBE and the design Sn density ranges from $10^{12}$ to $10^{14}$ cm$^{-2}$. Magnetotransport experiments were carried out at low temperatures in order to determine the relevant band structure parameters, such as the electron densities, the number of subbands and the mobilities. The electron densities determined from the Hall effect range from $1.7\times10^{12}$ up to $8.4\times10^{13}$ cm$^{-2}$. At the highest electron density electron subbands at the L point become populated. In this case also the DX centres are populated, which results in persistent photoconductivity effects at low temperatures [8]. The fairly low mobilities, as determined from Hall-effect measurements ($\mu_H \sim$ 1000 cm$^2$/Vs), dictate the need for high magnetic fields (B> 10 T) in order to observe the Shubnikov-de Haas effect. Therefore, parts of the experiments were carried out in pulsed fields up to 40 T generated by the High Magnetic Field Facility of the University of Amsterdam. In addition, the diamagnetic Shubnikov-de Haas effect was measured in order to determine the number of subbands [9,10]. The band structure was calculated self consistently and compared to the experimental data. The calculations show a typical thickness of the δ-layer of 160 Å.

## 2. Experimental

The Sn δ-doped GaAs structures were grown by MBE. First a buffer layer of *i*-GaAs was grown with thickness d= 240 nm on a semi-insulating GaAs(Cr) [001] substrate. Next a Sn layer was deposited in the presence of an arsenic flux at a substrate temperature of ~ 450 °C. Subsequently, a layer of *i*-GaAs (d= 40 nm) was grown, followed by a contact layer (d= 20 nm) of n-GaAs, with a Si concentration of 1.5x10$^{18}$ cm$^{-3}$, in order to suppress surface depletion. Several wavers (labelled N1-N7) with design Sn densities, $n_D$, ranging from 3.0x10$^{12}$ to 2.7x10$^{14}$ cm$^{-2}$ were grown (see Table I). From each waver a number of Hall bars and samples with van der Pauw geometry were prepared.

The longitudinal resistivity $\rho_{xx}(T)$ was measured in the temperature range 0.4-300 K. Temperatures above 4.2 K were obtained using a bath cryostat, while temperatures below 4.2 K were obtained with a $^3$He cryostat. The Hall resistance $\rho_{xy}(B)$ for a field perpendicular to the 2DEG was measured in the temperature range 0.4-12 K in stationary magnetic fields up to 10 T. The magnetotransport data were measured using a low-frequency ac-technique. In addition, magnetoresistance and Hall effect experiments were carried out in pulsed magnetic fields up to 38 T at T= 4.2 K. The samples were immersed in liquid helium to ensure stable temperatures. Shubnikov-de Haas data were taken with the pulse magnet in the free decay mode, after energising the magnet to the maximum field value. The total pulse duration amounts to 1 s. The high-field magnetotransport data were measured using a dc-technique with a typical excitation current I=1-10 μA. The high-field magnetoresistance was measured for a field perpendicular and parallel to the 2DEG.

## 3. Results

For all samples we find that the Hall resistance $\rho_{xy}(B)$ is a linear function of the magnetic field and does not vary with temperature in the range 0.4-12 K. The electron densities of samples N1-N7 determined from the Hall constant, $n_H$, range from 1.74x10$^{12}$ cm$^{-2}$ to 8.4x10$^{13}$ cm$^{-2}$ (see Table I). The electron density 8.4x10$^{13}$ cm$^{-2}$ measured for sample N7 is one of the highest values obtained by δ-doping. The mobility of the structures as determined from the Hall density and the zero-field resistivity at 4.2 K is fairly low and amounts to roughly 1000 cm$^2$/Vs (see Table I). At low temperatures (T< 4.2 K) a negative magnetoresistance was observed for all samples in low-magnetic fields (B< 0.2 T), which is attributed to the suppression of weak-localisation effects by the field. In the remaining part of the paper we predominantly focus on the samples N1, N2 and N7, which have quite different electron densities of 1.74x10$^{12}$, 14.5x10$^{12}$ and 8.4x10$^{13}$ cm$^{-2}$, respectively. These samples may be considered as exemplary for our study of the Sn δ-doped structures.

In Fig.1 we show some typical $\rho_{xx}(T)$ traces. The values of $\rho_{xx}$(4.2K) are 8.0, 1.35 and 0.17 kΩ for N1, N2 and N7, respectively, which shows that $\rho_{xx}$ decreases with increasing carrier concentration, as expected. At low values of $n_H$ $\rho_{xx}(T)$ shows a semiconducting-like behaviour (sample N1), while at the highest value of $n_H$ $\rho_{xx}(T)$ shows metallic behaviour (sample N7). Sample N2 presents an intermediate case, as its resistance shows a broad minimum around 175 K. These data show that at high carrier concentrations scattering at phonons is dominant, while at low carrier concentrations the scattering at ionised impurities dominates. Below ~10 K an extra correction is due to weak localisation with a logarithmic temperature dependence.

In Figs. 2a-4a we show some typical $\Delta\rho_{xx}=\rho_{xx}(B)-\rho_{xx}(0)$ traces, with B directed perpendicular to the δ-layer, for samples N1, N2 and N7, taken at T= 4.2 K. In all cases we observe pronounced Shubnikov-de Haas oscillations with several frequency components $F_i$, as clearly follows from the Fourier transforms (Figs. 2b-4b). We have verified the two-dimensional nature of these frequency components, by varying the angle between the magnetic field and the normal of the δ-layer. The expected behaviour $F(\theta)=F(0)/\cos\theta$ was observed (θ is the angle over which the δ-layer is rotated). In Figs. 5 and 6 we show $\rho_{xx}(B)$, with the magnetic field directed in the plane of the δ-layer for samples N2 and N7, respectively, at T= 4.2 K. The magnetoresistance was measured for B∥I and B⊥I, however no significant difference was observed. The oscillatory behaviour, as shown in Figs. 5 and 6, is due to the depopulation of subbands. In the next section we analyse and discuss the results.

## 4. Analysis and discussion

Inspecting the Fourier transforms of $\rho_{xx}(B)$, shown in Figs. 2b and 3b, we conclude that two and four frequency components, indicated by the peak labels (0-3), are present for samples N1 and N2, respectively. For sample N7 at least three, but possibly five frequency components are observed (Fig. 4b). Although, in general, some of the frequency components could be caused by higher harmonics and/or sum and difference frequencies, this is not the case here. For all samples the different frequencies can be associated with individual subbands, which is supported by the band structure calculations (see below).

The Shubnikov-de Haas frequency is related to the electron density by $n_{SdH}=2eF_i/h$, where e is the electron charge and h is Planck's constant. The factor 2 results from the spin degeneracy, which is not lifted. The values for $n_{SdH}$ of the individual subbands of samples N1, N2 and N7 are listed in Table II, where the label of the subband corresponds to the peak number in the Fourier transform. The sum of the values for $n_{SdH}$ of the different subbands ($\Sigma n_{SdH}$) is listed in Table I for samples N1-N7. In these multiple subband systems the comparison of $n_H$ and $n_{SdH}$ is not straight forward, notably because $n_H$ depends on the densities and mobilities of the different subbands. For sample N1, which has three occupied subbands, $n_H < n_{SdH}$, while for samples N2-N7 $n_H > n_{SdH}$. The latter behaviour has also been reported for heavily Si δ-doped structures [11].

As follows from the data in Table I $\Sigma n_{SdH}$ is not a monotonous function of the design doping density. We emphasise that this is not due to inhomogeneities of the wafers, as samples cut from different parts of the waver yield identical results (within 1%). The value of $\Sigma n_{SdH}$ is within a few percent equal to the doping density for samples N1 and N2, while for samples N3-N7 $\Sigma n_{SdH}$ is about a factor 5-10 lower than the design doping density. The difference between $n_D$ and $\Sigma n_{SdH}$ becomes significant near $n_D=1\times10^{13}$ cm$^{-2}$. We attribute this distinct difference to the formation of 3D tin islands at a doping density larger than $\sim 1\times10^{13}$ cm$^{-2}$. It is known that Sn may form 3D islands, depending on doping density and growth temperature [12]. The ability of Sn to aggregate into 3D islands has a strong influence on the mobilities of the electron subbands. This additional scattering mechanism hampers the observation of the electron subbands and therefore $\Sigma n_{SdH}$ is smaller than the doping density. One of our most important findings is that the electron density does not saturate at large design doping densities, as is observed in the case of Si δ-doped structures [13,14]. This indicates that the compensation mechanism, when Sn atoms are incorporated on As sites, is not significant.

In order to evaluate the electron quantum mobility $\mu_q^{SdH}$ of each subband, the corresponding SdH oscillation was separated using a digital filter [15] and a Dingle plot was made [16]. The resulting values obtained at T= 4.2 K are listed in Table II. A considerable variation in $\mu_q^{SdH}$ over the different subbands is observed, which is best illustrated by the data for sample N2, where $\mu_q^{SdH}$ ranges from 420 to 2040 cm$^2$/Vs. The relatively low quantum mobilities of the different subbands are reflected in part in the width of the Fourier peaks. For the lower subbands, the values of $\mu_q^{SdH}$ are similar to the values reported for Si δ-doped GaAs [17].

The quantum mobility is limited by the scattering at ionised Sn impurities [18-20]. Following Ref.18-20 we have calculated the quantum mobility, $\mu_q^{calc}$, including multiple subband scattering. The screening of the Coulomb scattering potential was taken into account within the random-phase approximation [19]. The results are listed in Table II, from which it follows that the calculated mobility increases with increasing subband number. For samples N1 and N2 the agreement between $\mu_q^{calc}$ and $\mu_q^{SdH}$ is quite good.

Next we present the band diagrams, wave functions and for each subband the electron concentration, that we have calculated by solving self-consistently the Schrödinger and Poisson equations [18]. The non-parabolicity of the conduction band [21] in the Γ-point and the exchange-correlation contribution [22] to the electrostatic potential were included in the calculations. The resulting band diagrams for samples N2 and N7 are shown in Figs. 7 and 8, respectively. The thickness of the δ-layer is used as an adjustable parameter in the calculations and amounts to 160 Å and 340Å for samples N2 and N7, respectively. This value is rather large compared to typical values for Si δ-layers [23-25] (20-100 Å), and is due to the large ability of Sn to segregate [4]. Another mechanism, which might result in a relatively large width, is the repulsive interaction between the ionised impurities [26]. This enhances the segregation, especially in heavily doped structures like N7. The calculated electron densities for each subband, $n_{calc}$, are listed in Table II. For samples N1 and N2 the agreement with the experimental values of $n_{SdH}$ is very good. The results for sample N7 will be discussed later in this section.

The oscillatory behaviour of the magnetoresistance for a magnetic field parallel to the δ-layer, as shown in Figs. 5 and 6 for samples N2 and N7, respectively, allows one to determine the number of occupied subbands [9,10]. This effect, which is sometimes termed the diamagnetic Shubnikov-de Haas effect, may serve as an useful tool to investigate two-dimensional multiple subband systems. This can be demonstrated by solving the Schrödinger equation with an in-plane magnetic field. Taking the field along the y-axis and the vector potential equal to **A**=(Bz,0,0) the Schrödinger equation for a wave function ψ with eigen-energy E reads [9]:

$$\left[\frac{\hbar^2 k_y^2}{2m^*} + \frac{(\hbar k_x + ezB)^2}{2m^*} - \frac{\hbar^2}{2m^*}\frac{\partial^2}{\partial z^2} + \Phi(z)\right]\Psi = E\Psi \qquad (1).$$

Here $k_x$ and $k_y$ are wavenumbers and m* is the effective electron mass. The potential Φ(z) is the sum of an electrostatic potential, determined from the Poisson equation, and the exchange-correlation potential [22]. The self-consistent solution can be written in the form $E=E_i(k_x)+\hbar^2 k_y^2/2m^*$, where i is the subband number and m*= $0.07 m_e$ ($m_e$ is the free electron mass). In Fig. 9 we show the dispersion relation $E_i(k_x)$ for B = 0 and B = 18T as calculated for sample N2. The subband energy levels shift towards the Fermi level with increasing magnetic field. In a field of 18 T the subbands i=4 and i=3 are depopulated. In Fig. 5 we show the total density of states (DOS), normalised at the zero field value, at the Fermi level for sample N2. With increasing magnetic field the DOS first increases, but drops when the subband level has crossed the Fermi level. This sharp singularity in the DOS indicates the depopulation field of the just emptied electron subband. In Table II we have listed the calculated depopulation fields, $B_\parallel^{calc}$, for samples N1, N2 and N7. When comparing the DOS and the oscillatory magnetoresistance a clear correspondence between the sharp discontinuities in the DOS and the minima in the first derivative versus magnetic field of the magnetoresistance $\partial\rho/\partial B$ are observed (see Fig. 5). Like in Ref. [9,10,27] we take the minimum values of $\partial\rho/\partial B$ as the experimental depopulation fields ($B_\parallel$). This is based on the analogy between minima in $\partial\rho/\partial B$ and maxima in $\partial\rho/\partial n$, where n is the electron density. In this case the maxima in $\partial\rho/\partial n$ mark the onset of the occupation of a new subband with increasing n, which is related to the DOS. The calculated values are in reasonable agreement with the experimental values, as determined from the minima in the first derivative of the magnetoresistance. The differences between the measured and calculated depopulation fields are attributed to mobility changes, which are not taken into account [28,29].

The measurements with the magnetic field parallel to the δ-layer were performed for two different field directions, with B∥I and B⊥I. The position of the minimum in $\partial\rho/\partial B$ does not depend on the direction between the in-plane magnetic field and the current. For thin Si δ-layers of approximately 20Å a shift of several Tesla was observed for the n=1 subband depopulation field when the sample was rotated from B⊥I to B∥I [10]. For thick Si δ-layers no current direction dependence was observed. Our measurements on Sn δ-layers are in correspondence with the latter observation.

The highest electron density was obtained for sample N7 ($n_H$=8.4×10$^{13}$ cm$^{-2}$). In the Fourier spectrum of sample N7 (Fig. 4b) the largest peak is observed at 207T with a shoulder at 232T. This broad peak with its shoulder corresponds to subbands 0-3 with electron densities 9.75-11.1×10$^{12}$ cm$^{-2}$, according to the bandstructure calculation (see Table II). These high electron subband densities are indicative for the multiple occupied subbands in this heavily doped sample. This is confirmed by the parallel magnetic field measurements, shown in Fig. 6. In a field of 38T six electron subbands are depopulated as illustrated by the insert. In the SdH effect and its Fourier transform (Fig. 4b) we do not observe the higher electron subbands (labelled 7-10 in Table II). This is because at the appropriate fields, where $\mu_q B$>1, the energy of the Fermi level is already close or below the lowest Landau level of these subbands. At low magnetic fields no subbands are observable due to the poor mobility.

From the high electron concentration of sample N7, we infer that the L conduction band is occupied (according to Ref. [13] this occurs at a concentration of ionised impurities greater than $n_D$=1.6×10$^{13}$ cm$^{-2}$). Assuming that the highest frequency in the Fourier spectrum (Fig. 4b) corresponds to the electron subband i=3 in the Γ point with concentration 1.0×10$^{13}$ cm$^{-2}$, then, according to our calculations, three subbands should be populated at the L point. The identification of the highest frequency with the i=3 subband at the Γ point gives the best agreement between the bandstructure calculations and experimental data. The measurements in perpendicular magnetic field do not show all the electron subbands. This complicates the calculation of the bandstructure. However combining the data obtained in $B_\parallel$ and $B_\perp$ improves the reliability of the bandstructure calculation.

For subbands at the L-point we have used the effective mass for quantisation in the $z$ direction $m_z = 0.11m_e$ and for the density of states $m^* = 0.38m_e$ (degeneracy $g_v=4$) [13]. The subband energy level structure in the L-point is calculated in a potential that is displaced by 290 meV from the Γ-point potential along the real-space co-ordinate. Good agreement with the experimental results is obtained when the width of δ-layer of ionised impurities is ~340Å. The band diagram of sample N7 is shown in Fig. 8. The unusual curvature of both the L- and Γ-potentials is connected with a complex total electron distribution consisting of occupied subband states at the Γ- and L-points. This is the first observation of the population of the L-point in δ-doped GaAs at low temperatures.

## 5. Conclusions

We have studied the magnetotransport properties of GaAs δ-doped structures with various Sn doping densities. Bandstructure calculations show that the thickness of the δ-layer in different samples is between 160Å and 340 Å. The maximum electron density determined by the Hall effect is $8.4\times10^{13}$ cm$^{-2}$ and strongly exceeds the maximum value for Si δ-doped structures. At such a high electron concentration the L conduction band is populated. The experimental electron subband densities and quantum mobilities are in good agreement with the calculated ones. The determination of the multiple subband structure was complemented by magnetoresistance measurements for a magnetic field parallel to the δ-layer. The measured values of the magnetic field, at which depopulation of subbands occurs, are in good agreement with calculated values of the depopulation fields.

## Acknowledgements

The work was part of the research program of the Dutch organization 'Stichting F.O.M.'. Support from the Russian Foundation for Basic Research (Grant N 97-02-17396) and the Dutch organization N.W.O., in the form of a Russian-Dutch research cooperation program, is gratefully acknowledged.

Figure captions

Fig. 1  Temperature dependence of the resistivity of samples N1, N2 and N7.

Fig. 2  Magnetoresistance of sample N1 (a) and the Fourier transform (b) at T=4.2K. The labels indicate the subband numbers.

Fig. 3  Magnetoresistance of sample N2 (a) and the Fourier transform (b) at T=4.2K. The labels indicate the subband numbers.

Fig. 4  Magnetoresistance of sample N7 (a) and the Fourier transform (b) at T=4.2K. The labels indicate the subband numbers in the Γ point. The broad peak with its shoulder corresponds to subbands 0-3, according to the bandstructure calculation.

Fig. 5  Magnetoresistance in parallel magnetic field for sample N2 (left axis) at T=4.2K. The calculated total density of states, normalised to the zero field value, at the Fermi energy for sample N2 (right axis).

Fig. 6  Magnetoresistance in parallel magnetic field for sample N7 at T=4.2K. The insert shows the first derivative dρ/dB.

Fig. 7  The calculated band structure for sample N2. The thick solid line denotes the potential well. The dashed lines indicate the energies $E_i$ of the electron subbands. The thin solid lines represent the electron wave functions for the different subbands. The shaded rectangle at the bottom of the figure gives the thickness of the doping layer.

Fig. 8  The calculated band structure for sample N7. The thick solid lines denote the potential wells for the Γ and the L points as indicated. The dashed lines indicate the energies $E_i$ of the electron subbands at the Γ point. The dotted lines indicate the energies $E_i$ of the electron subbands at the L point. The shaded rectangle at the bottom of the figure gives the thickness of the doping layer.

Fig. 9  Dispersion relation $E_i(k_x)$ for sample N2 in zero field (solid lines) and for a parallel magnetic field B = 18 T (dashed lines). In a field of 18 T subbands labelled 3 and 4 are depopulated.

Table I
The design doping density $n_D$, Hall concentration $n_H$, sum of the Shubnikov - de Haas concentrations $\Sigma n_{SdH}$ and the Hall mobility $\mu_H$ for GaAs($\delta$-Sn) samples N1-N7 at T=4.2K.

| sample # | $n_D$ ($10^{12}$ cm$^{-2}$) | $n_H$ ($10^{12}$ cm$^{-2}$) | $\Sigma n_{SdH}$ ($10^{12}$ cm$^{-2}$) | $\mu_H$ (cm$^2$/Vs) |
|---|---|---|---|---|
| N1 | 2.97 | 1.74 | 2.75 | 1530 |
| N2 | 8.90 | 3.59 | 8.73 | 1940 |
| N3 | 9.90 | 3.23 | 1.04 | 540 |
| N4 | 26.7 | 2.63 | 2.03 | 1080 |
| N5 | 29.7 | 10.4 | 6.15 | 1200 |
| N6 | 89.1 | 8.4 | 8.09 | 1150 |
| N7 | 267 | 84 | 44.8 | 1170 |

Table II

Experimental and calculated parameters for GaAs($\delta$-Sn) structures N1, N2 and N7 at T=4.2K. $n_{SdH}$ is the electron subband concentration, $\mu_q^{SdH}$ is the quantum mobility obtained from the Shubnikov - de Haas effect, $n_{cal}$ is the self-consistently calculated electron concentration, $\mu_q^{cal}$ is the calculated quantum mobility, $B_{||}$ is the experimentally determined depopulation field and $B_{||}^{cal}$ the calculated depopulation field.

| sample # | i subband number ($\Gamma$ point) | $n_{SdH}$ ($10^{12}$ cm$^{-2}$) | $n_{cal}$ ($10^{12}$ cm$^{-2}$) | $\mu_q^{SdH}$ (cm$^2$/Vs) | $\mu_q^{cal}$ (cm$^2$/Vs) | $B_{||}$ (T) | $B_{||}^{cal}$ (T) |
|---|---|---|---|---|---|---|---|
| N1 | 0 | 1.76 | 1.75 | 670 | 790 | - | - |
|    | 1 | 0.99 | 0.99 | 725 | 900 | 18.6 | 22.5 |
|    | 2 | -    | 0.30 | -   | 1120 | 4 | 8.0 |
| N2 | 0 | 3.80 | 3.75 | 420 | 650 | - | - |
|    | 1 | 2.40 | 2.68 | -   | 710 | - | 45.3 |
|    | 2 | 1.56 | 1.56 | 1000 | 1000 | 25.8 | 24.5 |
|    | 3 | 0.85 | 0.67 | 2040 | 1620 | 12.6 | 12.5 |
|    | 4 | -    | 0.06 | -   | 1170 | 4.4 | 3 |
| N7 | 0 | 11.2 | 11.06 | - | 217 | - | - |
|    | 1 | -    | 10.80 | - | 217 | - | - |
|    | 2 | -    | 10.38 | - | 218 | - | - |
|    | 3 | 10.0 | 9.75  | 690 | 220 | - | - |
|    | 4 | 9.09 | 8.87  | 830 | 225 | - | 43.9 |
|    | 5 | 8.17 | 7.84  | 890 | 236 | 32.0 | 34.1 |
|    | 6 | 6.3  | 6.68  | -   | 258 | 26.6 | 28.7 |
|    | 7 | -    | 5.36  | -   | 295 | 22.3 | 23.3 |
|    | 8 | -    | 3.91  | -   | 359 | 17.0 | 17.8 |
|    | 9 | -    | 2.49  | -   | 461 | 10.7 | 12.5 |
|    | 10 | -   | 1.32  | -   | 509 | 5.2 | 8.0 |